A MACHIAN DEFINITION OF PARTICLE MASS IN HIGHER-DIMENSIONAL GRAVITY

Paul S. Wesson

Dept. of Physics and Astronomy, University of Waterloo, Waterloo, ON, N2L 3G1, Canada.

Abstract: A new method involving the effective wave function is used to define the mass of a particle in a standard five-dimensional extension of general relativity. The mass is inversely proportional to the magnitude of the scalar field of the extra dimension. Since the scalar field is global and depends on the coordinates, this definition for particle mass agrees with Mach's Principle.

A MACHIAN DEFINITION OF PARTICLE MASS IN HIGHER-DIMENSIONAL GRAVITY

1. Introduction

The definition of the mass *m* of a particle in dimensionally-extended versions of general relativity is notoriously difficult. While the parameter *m* appears in many technical equations, which provide a context for it, at a basic level mass is a concept like space or time, all of which are to a certain extent vague in nature. In the present account I have a limited but useful aim. Namely to use a five-dimensional version of general relativity to follow a new route involving the wave function to a definition of the (inertial) mass of a particle. This definition will turn out to agree with Mach's Principle, in the sense that the mass of a local particle depends on the magnitude of a global scalar field.

The use of five dimensions is not arbitrary. In four-dimensional Einstein theory the mass is a given parameter with no deeper rationale. (Even so, it is a subtle quantity with several inequivalent definitions for the mass of a gravitating object.) By contrast, in modern versions of non-compactified Kaluza-Klein theory, mass plays a central role. But while Membrane theory and Space-Time-Matter theory are in agreement with observations, previous attempts to define the mass of a particle have proven to be in contradiction with both each other and with experiment (see below). Theories with more than five dimensions are not expected to be any easier to deal with, though they may someday yield a truly unified account of gravity and the interactions of particles.

In the next section, a brief survey is given of previous attempts to define particle mass in 5D relativity. Then the so-called canonical 5D metric is outlined, since it yields an effective wave function from which the mass can be obtained when the spacetime is an Einstein vacuum.

To describe a more realistic situation, the canonical metric may be generalized by introducing a global scalar field which is a function of the coordinates. This yields a definition for the mass of a particle as measured locally which depends inversely on the magnitude of the scalar field. Mach's Principle, wherein local mass depends on the rest of the universe, is thereby given a precise meaning.

Nomenclature is standard throughout. Upper-case letters $A$, $B$ etc. run 0, 123, 4 for time, space and the extra coordinate ($x^4 \equiv y$). Lower-case letters $\alpha, \beta$ etc. run 0, 123 for spacetime. To aid physical understanding at some places, conventional symbols are used for the gravitational constant $G$, Planck's constant $h$ and the speed of light $c$.

2.  Particle Mass in 4D and 5D Relativity

Early attempts to explain the origin and nature of rest mass were made in the 1960s and 1970s using classical 4D theory. A well-known example is the Brans-Dicke theory, which was motivated by Mach's Principle and added a scalar field to the conventional tensor interaction in order to describe gravity [1]. Later Dirac, Hoyle and Narlikar and others proposed versions of general relativity that were not only covariant under a change of coordinates but also invariant under a change of scale, this being accomplished through the addition of a scalar gauge function which had the effect of making the masses of particles variable [2,3]. None of these early theories gained widespread acceptance. However, Dirac in an even earlier paper had shown how a 5D embedding of the 4D de Sitter solution could be used to classify the properties of particles [4], an insight that was to resurface in other guises in more modern times.

In 1984, it was realized by Wesson and others that scale invariance in 4D was usually broken by the presence of constant particle masses in the relevant equations, something which could be avoided by introducing a fifth dimension related to rest mass [5, 6]. This also applies to the concept of supersymmetry as it is applied to particles, which should therefore be more logically formulated in 5D. The original 5D mass theory did not make significant use of the scalar field, which finds a natural place in 5D as the extra diagonal component of the metric tensor. Instead, the mass of a particle was taken to be proportional to the extra coordinate, $x^4 = y$. This primitive theory was superseded in 1992 by a more sophisticated approach, in which all matter in 4D (including particles) was induced by the fifth dimension. The main formulation for this was the algebraic proof that the 5D apparently empty Ricci equations $R_{AB} = 0$ contained the 4D Einstein equations $G_{\alpha\beta} = 8\pi T_{\alpha\beta}$ *with sources* [7]. This remarkable result was later realized to follow from an old embedding theorem of Campbell [8]. In the meantime, the induced-matter approach was applied to a class of 5D Friedmann-Robertson-Walker cosmological models originally discovered by Ponce de Leon [9], and found to give back all of the conventional matter properties and dynamics[10]. The new approach, which came to be known as Space-Time-Matter theory, also yielded the physical properties of the class of 5D objects known as solitons [11]. What might be termed the initial stage of the development of a 5D theory of matter and particles was effectively concluded in 1994, when Mashhoon and coworkers showed that there was only *one* coordinate frame in which the mass of a particle could be represented just by the extra coordinate, this being what came to be known as the 5D canonical metric [12]. I will take up this subject again below, because the 5D canonical metric also provides an embedding for all solutions of the 4D Einstein equations which are empty of ordinary matter but contain vacuum energy measured by a cosmological constant.

There have been many developments of 5D theory in recent years. A major one was the introduction in 1998 of Membrane theory, in which the 5D bulk manifold is split by a singular hypersurface about which the interactions of particles are concentrated [13, 14]. The main motivation for this is to explain the relatively great strengths of the other interactions compared to gravity, and the reason why the observed particles have masses much less than the theoretically-expected Planck value. Another major direction of research, closer to the current topic, was an effort to find an expression for the rest mass of a particle in a *general* 5D metric. In the aforementioned canonical metric, the extra coordinate appears as a prefactor on the 4D interval, via $(y/L)ds$ where $L$ is a constant length. The combination of parameters here resembles the conventional action for a particle of rest mass $m$, namely $mds$. Accordingly, in the original approach of Mashhoon et al [12], the identification was suggested $y = Gm/c^2$, which is dimensionally acceptable and uses the Schwarzschild radius of the particle to measure its gravitational mass. However, this inevitably makes the particle mass variable, since the equations of motion for the theory yield in general a path with $y = y(s)$. Another option presents itself, though. That is to make the identification $L = h/mc$, which is also dimensionally acceptable and uses the Compton wavelength of the particle to measure its inertial mass. The dynamics of the theory shows that $1/L \equiv (1/y)(dy/ds)$ is a constant of the motion, so a particle's inertial mass does not vary. The second alternative has seemed to some workers to be the one indicated by observations. But the first identification would also be acceptable if a particle's mass only varied slowly over cosmological timescales, and is also more in line with Mach's Principle [15]. These considerations, it should be noted, do not take into account constraints set by the Equivalence Principle. However, this can be viewed as a kind of symmetry of the 5D canonical metric, where the change between gravitational and inertial measures of the mass can be implemented by the coordinate transformation $y \to L^2/y$. The use of different coordinates in this way has been discussed by Wesson [16]. He has also pointed out that the alternative identification of *m* with the velocity $dl/ds$ in the extra dimension, as suggested by some authors, is really the first alternative noted above, because the existence of the constant *L* means that $dy/ds = y/L$, so the two options are dynamically equivalent. In summary, it should be reiterated that the 5D canonical metric is for all its usefulness only a special case, and a *general* definition for the mass of a particle must necessarily be more complicated that hitherto discussed.

An impressive body of work related to our problem has been carried out by Ponce de Leon [17-21]. He has studied the 5D equations of motion (geodesic and otherwise), the Hamilton-Jacobi method and the principle of least action. The last approach appears to be the broadest, and leads to the inference that in general the mass *m* varies along the particle's path [21]. This agrees with the results of other more restricted investigations, which are referenced and discussed by Ponce de Leon. It should be noted, however, that results from different approaches are not always concordant. Another valuable inference from the studies of Ponce de Leon, in addition to the variability of *m*, is that physically-acceptable motion in 4D follows from 5D paths which may be *null*. This agrees with previous work by Wesson [16], who also inquired if quantum effects in 4D may arise from paths which are null in 5D [22-24]. This is an important subject, to which we will return below.

In 4D general relativity, the motion of a test particle comes primarily from the metric, usually via the geodesic equation; while the properties of matter of the medium through which the particle moves come primarily from the field equations, usually via the assumption of a perfect fluid. The same plan also applies to 5D relativity. In what follows, we shall not be much concerned with field equations, but for completeness the reduction from 5D to 4D is given in compact form in the Appendix. In passing, it can be noted that by combining the equations found there and taking the trace of the effective energy-momentum tensor, it transpires that the density and pressure of any perfect fluid obeys the relation

$$8\pi|\rho - 3p| = \frac{1}{4\Phi^2}\left[g^{\mu\nu}{}_{,4}\, g_{\mu\nu,4} + (g^{\mu\nu}g_{\mu\nu,4})^2\right] \ .$$

Clearly the properties of matter scale as $1/\Phi^2$, where $\Phi(x^\gamma, y)$ is the scalar potential (a comma denotes the partial derivative with respect to $x^4 = y$). This relation is characteristic of the Space-Time-Matter approach, and because $\Phi$ plays such a central role in the theory we expect it also to figure in the dynamics. The aim in what follows is to derive a general expression for the inertial rest mass $m(x^\gamma, y)$ of a test particle.

As mentioned above, some insight to this problem can be obtained by considering the 5D canonical metric:

$$dS^2 = (y/L)^2 ds^2 + \varepsilon dy^2 \ . \tag{1}$$

In this so-called pure form, the embedded 4D interval $ds^2 = g_{\alpha\beta}(x^\gamma)dx^\alpha dx^\beta$ depends only on the spacetime coordinates. The first part of the 5D metric therefore depends on the extra coordinate only via the prefactor, which is chosen by analogy with the synchronous frame of standard cosmology. Using (1) with the field equations shows that it describes a vacuum spacetime. It has vacuum energy only, with the conventional equation of state $p = -\rho = -\Lambda/8\pi$. The cosmological constant can be positive or negative, depending on whether the extra dimension is spacelike ($\varepsilon = -1$) or timelike ($\varepsilon = +1$). Both are allowed, and $\varepsilon = +1$ in (1) does not lead to closed timelike paths because $x^4 = y$ does not have the nature of a time. (Also, the signature is tradi-tionally taken to be $+---+$ in Membrane theory.) Let us take $\varepsilon=+1$ to be definite. Further, we adopt as a postulate the definition $dS^2 \equiv 0$ for 5D causality, which is known to include the 4D conditions $ds^2 = 0$ and $ds^2 > 0$ for photons and massive particles [16, 21-24]. Then (1) gives

$$y = y_* e^{\pm is/L} \quad , \tag{2}$$

where $y_*$ is a fiducial constant. This is identical in form to the wave function of (old) quantum theory, provided $L$ is identified with the Compton wavelength $h/mc$ of the particle that accompanies the wave.

We infer that the wave of wave mechanics is an oscillation of the extra coordinate across spacetime, as viewed from the 5D perspective. This inference is confirmed by two studies: firstly, (2) in conjunction with the extra component of the 5D geodesic yields the Klein-Gordon wave equation for a relativistic particle [22]; secondly, (2) in conjunction with the de Sitter solution for the 4D part of (1) yields oscillations with the same properties as the de Broglie waves of standard wave mechanics [24]. More physics could be derived from (1), but for our purposes it should be recalled that it is a very restricted metric, notably in having $g_{\alpha\beta} = g_{\alpha\beta}(x^\gamma$ only) and $|g_{44}| = 1$. It is also intrinsically restricted to describing neutral particles, because the metric coefficients $g_{4\alpha}$ which normally correspond to the four electromagnetic potentials in Kaluza-Klein theories are absent.

To proceed in a tractable manner, we consider a neutral spinless particle (the quanta of the scalar field are in any case spin-0), but remove the other restrictions applied above. The 5D metric is now

$$dS^2 = (y/L)^2 g_{\alpha\beta}(x^\gamma, y) dx^\alpha dx^\beta + \Phi^2(x^\gamma, y) dy^2 \quad . \tag{3}$$

This form is actually completely general. It is true that it lacks the electromagnetic potentials, but these could in any case be removed by employing four of the available five coordinate degrees of freedom. (If they existed, the electromagnetic effects would then be concealed in the gravitational sector.) The synchronous prefactor $(y/L)^2$ remains, but as long as the rest of the 4D metric is allowed to depend on $x^4 = y$ this does not restrict the problem in a fundamental way. And the scalar field is now included in a general form, allowing of a range of background fluids through which the particle can move. (Recall the relation between the properties of matter and the scalar field noted before.) Putting $dS^2 = 0$ to define the null path, and partly integrating, (3) gives

$$y = y_* \exp[\pm i \int (ds/L\Phi)] \quad . \tag{4}$$

This replaces the special result (2), and includes it.

To make contact with conventional theory, let us consider the action of a particle integrated along its 4D trajectory and the corresponding wave function. Since the mass may be variable, it has to be kept inside the integral, so the standard forms are:

$$\int p_x dx^\alpha = \int m u_\alpha dx^\alpha = \int m ds$$

$$\psi = \psi_* \exp[\pm i \int m ds] \quad , \tag{5}$$

where the 4-velocities ($u^\alpha \equiv dx^\alpha/ds$) and the 4-momenta ($p^\alpha \equiv mu^\alpha$) are defined as usual. As before, we now compare the 5D and 4D expressions, or (4) and (5). The result is

$$m = 1/L\Phi \quad . \tag{6}$$

This is remarkable for its simplicity.

Some comments are in order about (6) and its application. (a) To evaluate (6), it is still necessary to determine $\Phi = \Phi(x^\gamma, y)$ from the 5D field equations. (b) For known solutions [16], it may be confirmed that the behaviour of $m$ in (6) is physically reasonable. (c) The canonical case (1) treated previously is included in (6) with $\Phi$ = constant, so a particle moving through an Einstein vacuum has constant mass. (d) The result (6) has the same dependency on $\Phi$ as one of the behaviours found by Ponce de Leon [21], though he used a classical approach while the one here uses the wave function.

It is really not surprising that metrics (1) and (3) lead to wave-like behaviour in the extra coordinate, since a related metric appears to mimic de Broglie waves [24]. This leads to the question whether such metrics can account for quantization. To inquire, it is instructive to generalize the pure-canonical metric (1) by adding a shift to the extra coordinate, via $y \rightarrow (y - y_0)$ where $y_0$ is a constant. This change has already been shown to lead to the appearance of a hypersurface where the energy density of the vacuum diverges [25]. This is similar to the symmetric, singular surface which is introduced "by hand" to Membrane theory [26]. However, in Space-Time-Matter theory this surface is a consequence of the geometry, and is not symmetric in the vacuum energy and is not truly singular (waves can cross it). To inquire further, we introduce a shift $y_0$ to the pure-canonical metric (1), so that the null-wave (2) acquires a new locus, and then split the metric near $y_0$ into two parts whose magnitudes are equal because of the null-path condition. The three relations expressing these steps are as follows:

$$dS^2 = \left(\frac{y-y_0}{L}\right)^2 ds^2 + dy^2 \tag{7}$$

$$y = y_0 + y_* e^{\pm is/L} \tag{8}$$

$$\left|\frac{ds}{L}\right| = \left|\frac{dy}{y-y_0}\right| . \tag{9}$$

The last of these is of particular interest. For as the membrane-like surface at $y_0$ is approached, $y \to y_0$ as $dy \to 0$ so the right-hand side of (9) tends to one. (This depends on the existence of the membrane at $y_0$, for without it the right-hand side is merely $dy/y$; see ref. 26 for a review of membranes and quantum-like consequences of other higher-dimensional theories of gravity.) The left-hand side of (9) also tends to one, so

$$ds \to L \text{ or } mc\,ds \to h \quad , \tag{10}$$

where $L = h/mc$ has been used from before. That is, there is quantization near $y = y_0$ with the usual rule.

## 3. Conclusion

Attempts to define the rest mass of a particle in theories of Kaluza-Klein type have a long and conflicted history. This is partly because a 5D metric offers several plausible (but inequivalent) approaches to the problem, and partly because the mass is expected in any case to be variable. As far as I am aware, all previous attempts have employed classical approaches (geodesic, Hamilton-Jacobi and least action). By contrast, the method adopted in the present ac-count has a lot in common with old quantum theory (wave mechanics). This approach is suggested by recent work which shows that 5D metrics of the canonical type lead to quantum consequences, notably de Broglie waves [22-24]. The 5D pure-canonical metric (with $\Phi$=constant) is known to embed all solutions of Einstein's equations which are empty of ordinary matter but possesses vacuum energy (measured by the cosmological constant). A broader form of the canonical metric leads straightforwardly to a general definition for the mass of a particle. This is $m=h/cL\Phi$, in conventional units, where $L$ is the length scale which occurs in conjunction with the extra coordinate and $\Phi$ is the scalar field of the theory. This scalar field is global, and depends in general on the spacetime coordinates and the extra coordinate (though the latter dependency can in simple cases be removed by a coordinate transformation). Thus the mass of a particle is in a sense dependent on its surroundings. To this extent, the definition of (inertial) mass given here is Machian.

This suggests an application to cosmological and astrophysical solutions of the 5D field equations, many of which are known [e.g. 9-11]. However, another field for application of the mass formula given here is particle physics. A traditional way to approach particle masses in 5D theory has been to consider a sum of components of the scalar field, but this frequently led to a tower of states with little connection to real particles. The definition $m \sim 1/\Phi$ found in the present work suggests that a better approach might be to look for solutions where $\Phi(x^\gamma, y)$ is relatively complicated and so more realistic. To extend the scheme discussed here to include electric charge and spin would involve adding the relevant off-diagonal components to the metric. And to incorporate the internal symmetries displayed by real particles would presumably involve adding more dimensions.

Another subject for further study concerns waves and quantization. The present account has shown that 5D metrics of canonical type lead to wave behaviour if the signature is appropriately chosen, and quantization according to the usual rule can occur if there is a membrane-like hypersurface present. This does not, of course, represent an approach to a quantum theory of gravity, but rather a demonstration of quantized waves within a classical theory. However, it is an interesting development, and may repay further investigation.

Acknowledgements

Thanks for comments go to J.M. Overduin and other members of the Space-Time-Matter group (5Dstm.org).

Appendix: The 5D Field Equations in 4D Form

It is convenient to use the following metric:

$$dS^2 = g_{\alpha\beta}(x^\gamma, y)dx^\alpha dx^\beta + \varepsilon \Phi^2(x^\gamma, y)dy^2 \quad . \tag{A1}$$

With this metric, the field equations $R_{AB} = 0$ can be conveniently grouped into sets of 10 (tensor), 4 (vector) and 1 (scalar), thus:

$$G_{\alpha\beta} = 8\pi T_{\alpha\beta}$$

$$8\pi T_{\alpha\beta} \equiv \frac{\Phi_{,\alpha;\beta}}{\Phi} - \frac{\varepsilon}{2\Phi^2}\left\{\frac{\Phi_{,4}g_{\alpha\beta,4}}{\Phi} - g_{\alpha\beta,44} + g^{\lambda\mu}g_{\alpha\lambda,4}g_{\beta\mu,4}\right.$$

$$-\frac{g^{\mu\nu}g_{\mu\nu,4}g_{\alpha\beta,4}}{2}+\frac{g_{\alpha\beta}}{4}\left[g^{\mu\nu}{}_{,4}g_{\mu\nu,4}+\left(g^{\mu\nu}g_{\mu\nu,4}\right)^2\right]\right\} \quad . \tag{A2}$$

$$P^\beta_{\alpha,\beta}=0$$

$$P^\beta_\alpha \equiv \frac{1}{2\Phi}\left(g^{\beta\sigma}g_{\sigma\alpha,4}-\delta^\beta_\alpha g^{\mu\nu}g_{\mu\nu,4}\right) \quad . \tag{A3}$$

$$\Box\Phi=-\frac{\varepsilon}{2\Phi}\left[\frac{g^{\lambda\beta}{}_{,4}g_{\lambda\beta,4}}{2}+g^{\lambda\beta}g_{\lambda\beta,44}-\frac{\Phi_{,4}g^{\lambda\beta}g_{\lambda\beta,4}}{\Phi}\right]$$

$$\Box\Phi \equiv g^{\alpha\beta}\Phi_{,\alpha,\beta} \quad . \tag{A4}$$

Here a comma denotes the partial derivative, and a semicolon denotes the standard (4D) covariant derivative. These equations, which were first derived by Wesson and Ponce de Leon [7], are algebraically general and can be applied to any physical problem where gravitational and scalar fields are dominant.